\documentclass[%
reprint,
superscriptaddress,
aip, 
amsmath,amssymb,
]{revtex4-2}

\usepackage{graphicx}
\usepackage{dcolumn}
\usepackage{bm}

\begin{document}

\preprint{APS/123-QED}

\bibliographystyle{naturemag}

\title{Role of nonlocal heat transport on the laser ablative Rayleigh–Taylor instability}

\author{Z. H. Chen}
\affiliation{College of Science, National University of Defense Technology, 410073 Changsha, China}%
 
\author{X. H. Yang}\thanks{xhyang@nudt.edu.cn}
\affiliation{College of Science, National University of Defense Technology, 410073 Changsha, China}
\affiliation{Collaborative Innovation Centre of IFSA, Shanghai Jiao Tong University, 200240 Shanghai, China}

\author{G. B. Zhang}
\affiliation{College of Science, National University of Defense Technology, 410073 Changsha, China}%

\author{Y. Y. Ma}
\affiliation{Collaborative Innovation Centre of IFSA, Shanghai Jiao Tong University, 200240 Shanghai, China}%
\affiliation{College of Advanced Interdisciplinary Studies, National University of Defense Technology, 410073 Changsha, China}%

\author{R. Yan}
\affiliation{Collaborative Innovation Centre of IFSA, Shanghai Jiao Tong University, 200240 Shanghai, China}%
\affiliation{Department of Modern Mechanics, University of Science and Technology of China,  230026 Hefei, China}%

\author{H. Xu}
\affiliation{College of Science, National University of Defense Technology, 410073 Changsha, China}%
\affiliation{Collaborative Innovation Centre of IFSA, Shanghai Jiao Tong University, 200240 Shanghai, China}%

\author{Z. M. Sheng}
\affiliation{Collaborative Innovation Centre of IFSA, Shanghai Jiao Tong University, 200240 Shanghai, China}%
\affiliation{Key Laboratory for Laser Plasmas (Ministry of Education), School of Physics and Astronomy, Shanghai Jiao Tong University, 200240 Shanghai, China}%

\author{F. Q. Shao}
\affiliation{College of Science, National University of Defense Technology, 410073 Changsha, China}%

\author{J. Zhang}\thanks{jzhang1@sjtu.edu.cn}
\affiliation{Collaborative Innovation Centre of IFSA, Shanghai Jiao Tong University, 200240 Shanghai, China}%
\affiliation{Key Laboratory for Laser Plasmas (Ministry of Education), School of Physics and Astronomy, Shanghai Jiao Tong University, 200240 Shanghai, China}%

\begin{abstract}
Ablative Rayleigh-Taylor instability (ARTI) and nonlocal heat transport are the critical problems in laser-driven inertial confinement fusion, while their coupling with each other is not completely understood yet. Here the ARTI in the presence of nonlocal heat transport is studied self-consistently for the first time theoretically and by using radiation hydrodynamic simulations. It is found that the nonlocal heat flux generated by the hot electron transport tends to attenuate the growth of instability, especially for short wavelength perturbations. A linear theory of the ARTI coupled with the nonlocal heat flux is developed, and a prominent stabilization of the ablation front via the nonlocal heat flux is found, in good agreement with numerical simulations. This effect becomes more significant as the laser intensity increases. Our results should have important references for the target designing for inertial confinement fusion.

\end{abstract}

\maketitle

\section{Introduction}
The electron thermal conduction is a key process in inertial confinement fusion (ICF) as it largely determines the implosion energy and the hydrodynamic instabilities on the ablation front. The classical Spitzer-Härm (SH) thermal conduction model becomes inadequate when the electron mean free path $\lambda_{ei}$ fails to meet the condition of being significantly smaller than the temperature gradient scale length ${L_T} = \left| {T/\nabla T} \right|$ , i.e., ${\lambda _{ei}}/{L_T} \ge 0.01$, where $T$ is the temperature\cite{lucianiNonlocalHeatTransport1983,bellNonSpitzerHeatFlow1985,moraNonlocalElectronTransport1994,gregoriEffectNonlocalTransport2004,henchenObservationNonlocalHeat2018,atzeniShockIgnitionThermonuclear2014}. The energy transport of electrons will be nonlocal, which will both limit the heat flux and preheat the interior of the ablation front. Heat flux inhibition has an impact on various aspects, including the corona temperature, laser energy absorption, ablation rate, ablation pressure, conduction zone length, and shock amplitude. On the other hand, the preheating leads to higher entropy and reduced compression in the unablated region. Additionally, it enhances the mass ablation rate while flattening the ablation front. It may have an important impact on the ablative Rayleigh-Taylor instability (ARTI). Up to now, the effects of electron nonlocal transport on the stability and symmetry of the ablation front have not yet been well-understood.

When the energy from laser irradiation or X-rays is absorbed by the shell for direct drive or indirect drive ICF, extensive ablation of the outer surface of the shell develops. From the perspective of the ablation front's reference frame, the ablation plasma accelerates the dense shell, ultimately giving rise to the occurrence of ARTI\cite{bodnerRayleighTaylorInstabilityLaserPellet1974,kullTheoryRayleighTaylorInstability1991,kilkennyReviewAblativeStabilization1994,pirizHydrodynamicInstabilitiesAblation2004,bettiBubbleAccelerationAblative2006,bettiInertialconfinementFusionLasers2016,zhouRayleighTaylorRichtmyer2017,zhangEnhancedEnergyCoupling2020,ramisAnalysisThreedimensionalEffects2019,matsuoEnhancementAblativeRayleighTaylor2021a,qiaoNovelTargetDesigns2021}. It is well known that the mass ablation leads to a reduction in the ARTI growth rate and provides complete stabilization for the perturbations with wavelengths shorter than a cutoff wavelength $\lambda_{cut}$. The linear growth rate of the ARTI can be well approximately by the modified Takabe's formula\cite{bettiGrowthRatesAblative1998}, $\gamma  = \alpha \sqrt {kg/\left( {1 + k{L_m}} \right)}  - \beta k{V_a}$, where $\alpha$ and $\beta$ are dimensionless constants depending on ablator materials, $k$ is the perturbation wavenumber, $L_m$ is the minimum density gradient scale length, $g$ is the acceleration, and $V_a$ is the ablation velocity. Further, the analytical solution for perturbed growth has been cleverly obtained by asymptotic matching techniques\cite{sanzSelfconsistentAnalyticalModel1994} and the WKB methods\cite{goncharovSelfconsistentStabilityAnalysis1996a,goncharovSelfconsistentStabilityAnalysis1996,bettiSelfconsistentCutoffWave1995}. It was found that the stabilizing mechanisms are attributed to vorticity convection\cite{kullTheoryRayleighTaylorInstability1991,kilkennyReviewAblativeStabilization1994} and dynamical overpressure\cite{pirizRayleighTaylorInstability1997,goncharovTheoryAblativeRichtmyerMeshkov1999}. It has been demonstrated through both simulations and experiments that the actual ARTI growth rates can be lower than predicted by the theories\cite{shigemoriMeasurementsRayleighTaylorGrowth1997,azechiDirectdriveHydrodynamicInstability1997,glendinningMeasurementDispersionCurve1997,hondaEffectsNonlocalElectron1999,sakaiyaAblativeRayleighTaylorInstability2002,smalyukRayleighTaylorGrowthStabilization2008,smalyukSystematicStudyRayleigh2008,liMitigationAblativeRayleigh2022}. This is usually attributed to the fact that the nonlocal effect can increase $V_a$ and $L_m$ via preheating, thus effectively suppressing the growth of the perturbation. The trend and cutoff wavelength of the growth rate curve can be roughly predicted using increasing $V_a$ and $L_m$. However, there is still a significant gap exists between the predicted growth rates and the experimental observations\cite{smalyukRayleighTaylorGrowthStabilization2008,smalyukSystematicStudyRayleigh2008}. This indicates that our comprehension of the impacts of electron nonlocal transport on ARTI, along with any potential stabilizing mechanisms, remains incomplete. There is not any theory so far on the ARTI growths with the electron nonlocal transport taken into account.

In this paper, we investigate the ARTI in planar laser-ablated foils with nonlocal heat transport considered self-consistently for the first time, where a nonlocal electron transport model known as the SNB model\cite{schurtzNonlocalElectronConduction2000} is applied. It is shown that the growth rate, when considering the local heat flux, aligns well with predictions from by the classical ARTI theory. However, if assuming that nonlocal effects increase $V_a$ and $L_m$ only, the theory fails to fully explain the stabilization of ARTI by the nonlocal transport found in the simulation. A linear theory coupling with the nonlocal heat flux is proposed, which is in good agreement with simulations. The observation highlights that the nonlocal heat flux exerts a more pronounced inhibitory effect on perturbations with shorter wavelengths, leading to an increase in the cutoff wavelength.

\section{Results and Discussions}
\subsection{Numerical Simulations}
\begin{figure}[htbp]\centering
	\includegraphics[width=8.5cm]{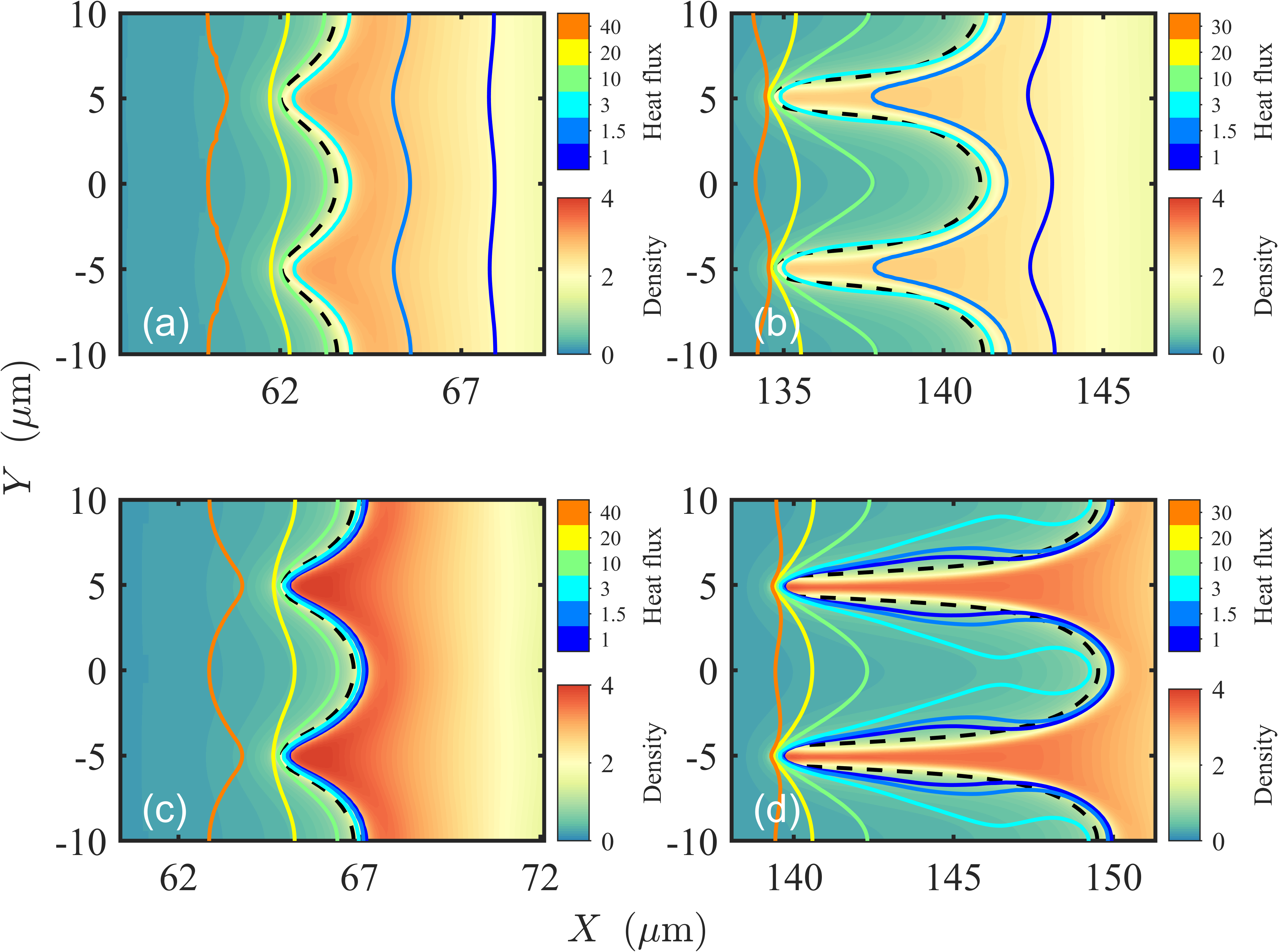}
	\caption{\label{fig1}The density profiles obtained with the SNB model [(a) and (b)] and the SH model [(c) and (d)] at time $t = 1\ \rm{ns}$ [(a) and (c)] and $t = 1.5\ \rm{ns}$ [(b) and (d)], where the density is in unit of $\rm{g/cm^3}$, the black dashed line signifies the ablation front and the irradiation originates from the left side. The transverse heat flux contours are overlaid on the density profiles in units of $10^{12}\ \rm{W/cm^2}$.}
\end{figure}

We have performed simulations of a single-mode ARTI by using the radiation hydrodynamic code FLASH, in which different heat conduction models can be taken, such as the SH model or the SNB model\cite{fryxellFLASHAdaptiveMesh2000,chenEffectNonlocalTransport2023}. A planar CH foil is irradiated by a $2\omega$ square-wave laser with peak intensity $4\times10^{14}\ \rm{W/cm^2}$ and pulse duration 2 ns. Initial cosine perturbations of $10\ \rm{\mu m}$ wavelength with an initial amplitude of $0.5\ \rm{\mu m}$ are set on the leading surface of the planar CH foil. The thickness of the target is $40\ \rm{\mu m}$. Under the irradiation of the laser on the left side, the target starts to accelerate with an average acceleration ${g^{SNB}} = 148.06\ \rm{\mu m/ns^2}$ and an average ablation velocity $V_a^{SNB} = 3.34\ \rm{\mu m/ns}$ when the SNB model is included. Figures \ref{fig1}(a) and \ref{fig1}(b) show the corresponding density profiles at time $t = 0.8$ and $1.5\ \rm{ns}$, respectively. The presence of transverse heat flux within the dense target is evident as a result of hot electron preheating. It increases the entropy of the target, decreases the compressive density and increases the ablation rate, thus inhibiting the growth of ARTI. The peak-to-valley heights at the two moments are $h_1^{SNB} = 1.56\ \rm{\mu m}$ and $h_2^{SNB} = 6.72\ \rm{\mu m}$, respectively. As a comparison, we have performed simulation with the classical SH heat conduction model under the same laser and target conditions. As shown in Figs. \ref{fig1}(c) and \ref{fig1}(d), the target moves with a higher average acceleration ${g^{SH}} = 150.44\ \rm{\mu m/ns^2}$ and a lower average ablation velocity $V_a^{SH} = 2.69\ \rm{\mu m/ns}$. The differences of the target acceleration are due to the slightly higher laser energy absorption of the SH model compared to the SNB model. In the meanwhile, the peak-to-valley heights obtained with the SH model at both moments are given by $h_1^{SH} = 2.19\ \rm{\mu m}$ and $h_2^{SH} = 10.47\ \rm{\mu m}$, respectively, larger than that with the SNB model.

\begin{figure}[htbp]\centering
	\includegraphics[width=9cm]{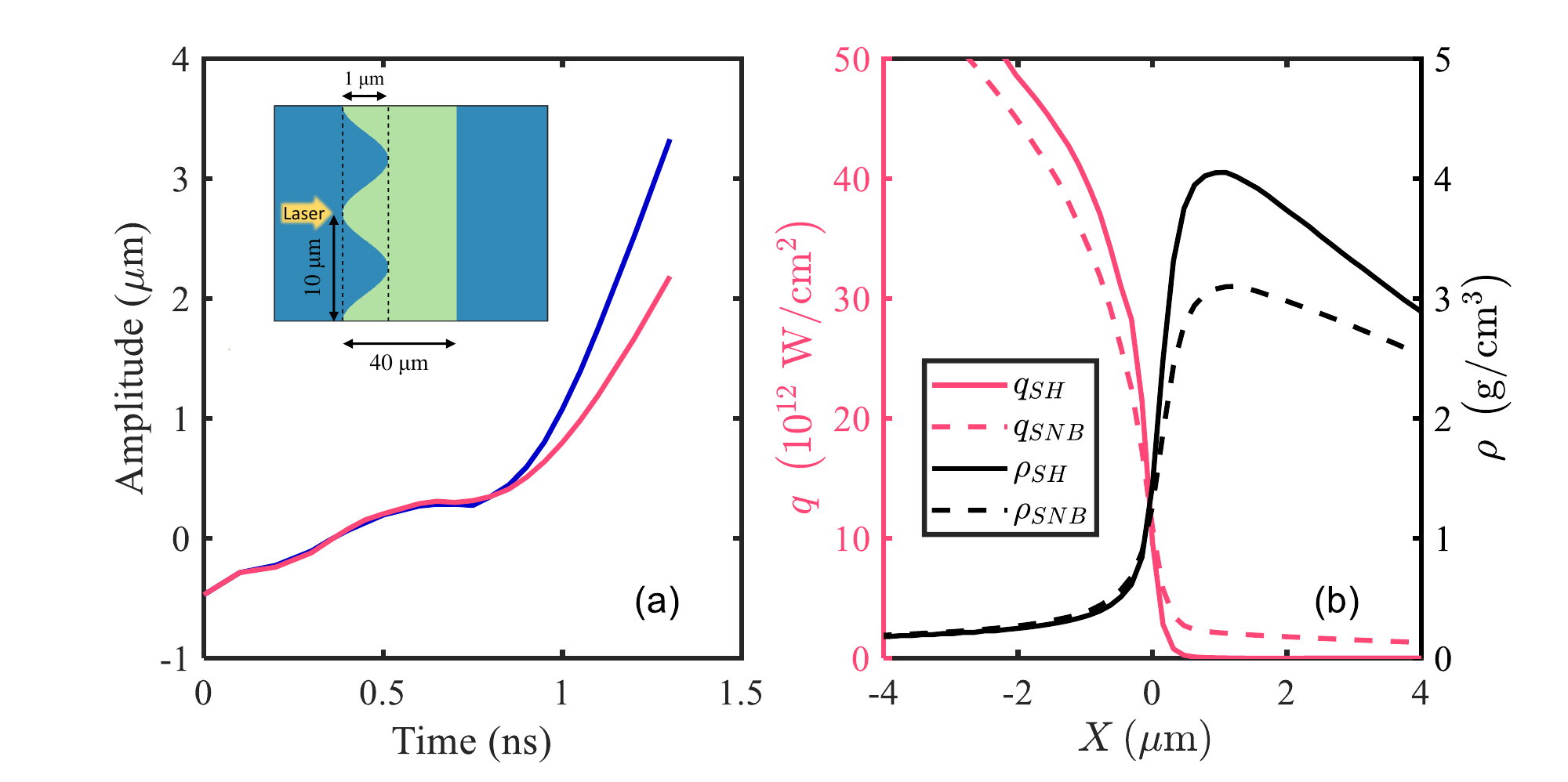}
	\caption{\label{fig2}(a) The evolution of perturbation amplitude over time reveals a phase reversal occurring at approximately $0.4\ \rm{ns}$. The inset shows the initial setup. (b) The profiles of heat flux and density around the perturbation peak along the acceleration direction at $1.0\ \rm{ns}$, with both models having their ablation fronts moved to $X = 0$.}
\end{figure}

Figure \ref{fig2}(a) shows the perturbation amplitude evolution in both cases, where a similar evolution in the Richtmyer–Meshkov instability (RMI) phase before $t = 0.7\ \rm{ns}$ are shown due to the similar laser energy absorption rate. The target starts to accelerate and the perturbation enters the RTI phase after $0.7\ \rm{ns}$. During the linear phase, which spans from $0.8$ ns to amplitudes reaching $0.1\sim0.2\lambda$, the growth rates for SNB and SH are $\gamma _{SNB}^{sim} = 4.20\ \rm{ns^{-1}}$ and $\gamma _{SH}^{sim} = 5.63\ \rm{ns^{-1}}$, respectively, where $\lambda$ is the perturbation wavelength. The profiles of density and transverse heat flux at the perturbation peak along the acceleration direction are presented in Fig. \ref{fig2}(b). The SNB model with preheating gives a lower peak density and larger ablation front characteristic length. Table \ref{tab:1} lists the linear growth rates derived from the simulations and predicted by Kull’s theory\cite{kullAblativeStabilizationIncompressible1986}, as well as the relevant parameters determining the growth rates. The Kull’s theory, henceforth called the classical theory, correctly predicts the linear growth rate for the classical SH model. However, it overestimates the growth rate even though that we apply the enhanced $V_a$ and $L_m$ from the nonlocal simulations. This suggests that a cautious approach should be taken when handling the hydrodynamic instability under the effects of nonlocal transport of electrons, as it cannot be adequately addressed solely by increasing $V_a$ and $L_m$.

\begin{table*}[htbp]
	\centering
	\caption{\label{tab:1}Linear ARTI growth rate parameters for direct-drive planar CH targets determined from FLASH simulations, where $R = {\rho _h}/{\rho _l}$ is the density ratio of heavy to light fluid, ${L_0} = {L_m}{\nu ^\nu }/{\left( {\nu  + 1} \right)^{\nu  + 1}}$ is the characteristic thickness of the ablation front, $\nu$ is the power index for thermal conduction, $\gamma _{sim}$ is the growth rate obtained by fitting the perturbation amplitude in the simulation, and $\gamma _{th}$ is the growth rate predicted from the theory\cite{kullAblativeStabilizationIncompressible1986}.}
	\begin{ruledtabular}
	\begin{tabular*}{\textwidth}{@{\extracolsep\fill}lcccccc}
		~ & ${V_a}{\rm{ }}\left( {{\rm{\mu m/ns}}} \right)$ & $g{\rm{ }}\left( {{\rm{\mu m/n}}{{\rm{s}}^{\rm{2}}}} \right)$ & ${L_0}{\rm{ }}\left( {{\rm{\mu m}}} \right)$ & $R$& ${\gamma _{sim}}{\rm{ }}\left( {{\rm{\mu }}{{\rm{m}}^{{\rm{ - 1}}}}} \right)$& ${\gamma _{th}}{\rm{ }}\left( {{\rm{\mu }}{{\rm{m}}^{{\rm{ - 1}}}}} \right)$ \\
		\hline
		SNB & 3.3670 & 148.06 & 0.0497 & 3.8846 & 4.20 & 4.99\\
		SH  & 2.6893 & 150.44 & 0.0369 & 4.2926 & 5.63 & 5.77 \\
		\end{tabular*}
	\end{ruledtabular}
\end{table*}

\subsection{Development of the Nonlocal Linear Theory}
To accurately predict the linear growth rate of ARTI, we have developed a linear theory that incorporates the nonlocal heat flux. Due to the complexity of the nonlocal multigroup diffusion equation, it is difficult to obtain an analytical solution for the ARTI growth rate. Therefore, this theory is based on the sharp boundary model (SBM)\cite{pirizHydrodynamicInstabilityAblation2001,sanzNonlinearTheoryAblative2002}, following the numerical approach akin to that employed by Kull et al.\cite{kullAblativeStabilizationIncompressible1986}. Utilizing the nonlocal multigroup diffusion equation proposed by Schurtz et al.\cite{schurtzNonlocalElectronConduction2000}, the energy equation can be written as $\nabla  \cdot \left( {{\boldsymbol{v}} + {\boldsymbol{Q}}/\left( {{c_p}\rho T} \right)} \right) = 0$ with the quasi-isobaric approximation, where ${\boldsymbol{v}}\left( {{v_x},{v_y}} \right)$ is the fluid velocity, $\rho$ is the density, $T$ is the temperature, $c_p$ is the specific heat capacity under constant pressure, ${\boldsymbol{Q}} =  - \kappa \nabla T - \sum\limits_g {{\lambda _g}\nabla {H_g}} /3$ is the total heat flux, $\kappa$ is the thermal conductivity, $\lambda_g$ is the group mean free path, and $H_g$ is the solution to the multigroup diffusion equations\cite{schurtzNonlocalElectronConduction2000}. The detailed derivation process is shown in the Methods.

\begin{figure}[htbp]\centering
	\includegraphics[width=9cm]{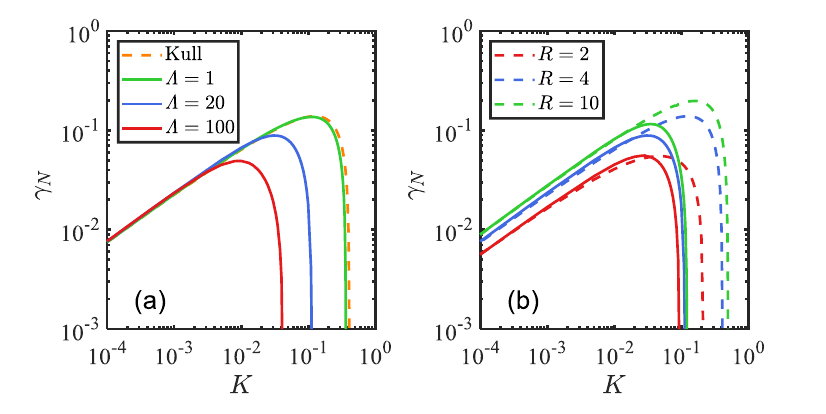}
	\caption{\label{fig3}(a) Growth rate curves for $\varLambda = 1, 20, 100$, $R = 4$, $S = 0.7$, and $\nu  = 2.5$. (b) Growth rate curves for $R = 2, 4, 10$, $\varLambda = 20$, $S = 0.7$ and $\nu  = 2.5$, where the solid line corresponds the nonlocal theory and the dashed line represents the classical theory.}
\end{figure}

The parameter $\varLambda = k_{c1}\lambda_1$, initially introduced in the nonlocal linear theory, is related to the electron characteristic free path $\lambda_1$ and can be used to estimate the strength of the preheating, where ${k_{c1}} = g/v_1^2$ , $v_1$ is the ablation velocity. Fig \ref{fig3}(a) shows the dimensionless growth rate ${\gamma _N} = \gamma /\left( {{k_{c1}}{v_1}} \right)$ for $\varLambda = 1, 20, 100$, $R = 4$, $S = 0.7$, and $\nu  = 2.5$, where $K = k/{k_{c1}}$ is the dimensionless wavenumber, $S = {k_{c1}}{\chi _1}/{v_1}$ is the reciprocal of Froude number, $R$ is the density ratio of the heavy fluid to the light fluid. This is analogous to increasing $\lambda_1$ while keeping the ablation velocity $V_a$, the characteristic length of the ablation front $L_0$ and the acceleration $g$ unchanged. At $\varLambda = 1$, the nonlocal theory essentially replicates the findings of the classical theoretical results. The growth rate curve has a maximum and cuts off at large wavenumbers due to ablative effects. As $\varLambda$ gradually increases, the cutoff wavenumber decreases. This implies that the nonlocal heat flux constitutes one of the important stabilizing factors for ARTI. It exerts varying degrees of suppression on perturbations of different wavelengths, displaying greater effectiveness in dampening perturbations with larger wavenumbers. Figure \ref{fig3}(b) shows the growth rate from the nonlocal and classical theories for $R = 2, 4, 10$. As $R$ increases, the growth rate curves have a higher peak and a larger cutoff wavenumber, underscoring the heightened influence of the nonlocal heat flux on the cutoff wavenumber.

\begin{figure}[htbp]\centering
	\includegraphics[width=9cm]{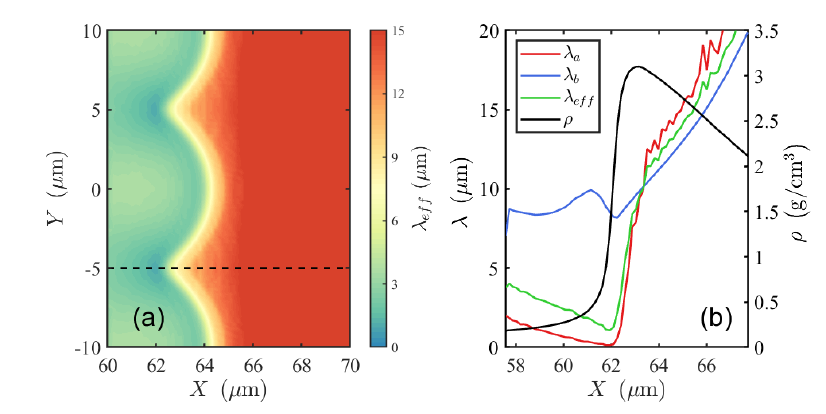}
	\caption{\label{fig4}(a) Spatial distribution of the effective free path $\lambda_{eff}$ for a $10\ \rm{\mu m}$ perturbation wavelength at $1\ \rm{ns}$. (b) Distribution of the effective free path $\lambda_{eff}$ along the direction of the dashed line in (a).}
\end{figure}

Figure \ref{fig4}(a) shows the distribution of $\lambda_{eff}$ for the case with a perturbation wavelength of $10\ \rm{\mu m}$ at $t = 1\ \rm{ns}$, where ${\lambda _{eff}} = \sqrt {{\lambda _a}{\lambda _b}}$, ${\lambda _a} = \sum\limits_g {{H_g}} /\left( {\sum\limits_g {{H_g}/{\lambda _g}} } \right)$ and ${\lambda _b} = \sum\limits_g {{\lambda _g}\nabla {H_g}} /\sum\limits_g {\nabla {H_g}} $. Along the direction of the perturbation peak, we obtain the one-dimensional profile of $\lambda_{eff}$, i.e., Fig. \ref{fig4}(b). It is worth noted that $\lambda_{eff}$ exhibits a minimum value in proximity to the ablation front. The growth rate of the nonlocal theory, $\gamma _{th}^{nonlocal} = 4.22\ \rm{ns^{-1}}$, is determined by inserting the average $\lambda_1$ observed during the linear phase into the nonlocal dispersion relation and $\lambda_1$ is taken as the value of $\lambda_{eff}$ near the ablation front. The theoretical growth rate is in good agreement with the that of the simulation, i.e., ${\gamma _{SNB}} = 4.20\ \rm{ns^{-1}}$. It is found that the variation of $\lambda_{eff}$ is dominated by $\lambda_{a}$. For the dense compressed region, the contribution of $\boldsymbol{Q}_{SH}$ is neglected and the value of $\lambda_{eff}$ increases with increasing depth into the target. Since ${\lambda _a} = \sum\limits_g {{\lambda _g}\nabla  \cdot {\boldsymbol{Q}}_{SNB}^g} /\nabla  \cdot {{\boldsymbol{Q}}_{SNB}}$($\nabla  \cdot\boldsymbol{Q}^g_{SNB}={H_g}/{\lambda _g}$), a larger $\lambda_{eff}$ indicates that the hot electrons contribute more to the change of energy. For the outer side of the ablation front, $\lambda_{eff}$ also increases during the process away from the target. This phenomenon arises because $\boldsymbol{Q}_{SNB}$ is significantly smaller than $\boldsymbol{Q}_{SH}$ in this region, and the electron mean free path $\lambda_{ei}$ gradually increases outward. Consequently, $\lambda_{eff}$ serves as a gauge of the intensity of nonlocal effects within the ablation front.

\begin{figure}[htbp]\centering
	\includegraphics[width=9cm]{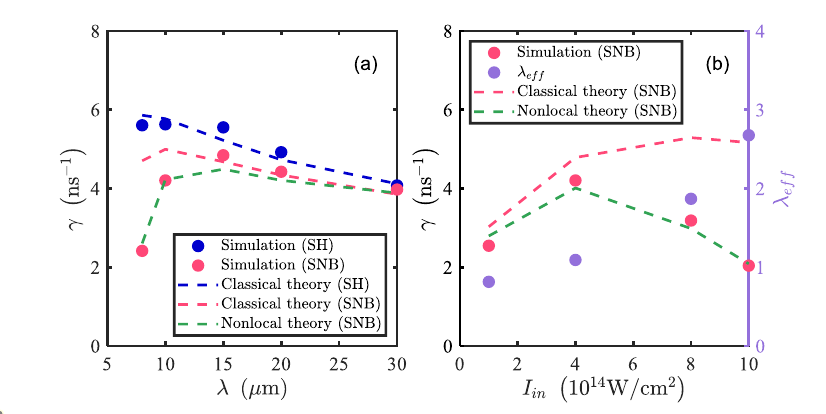}
	\caption{\label{fig5}(a) Linear growth rates of the ARTI with the two models, where the results for the classical and nonlocal theories are the growth rates obtained by substituting the hydrodynamic parameters from the simulations (SH or SNB model) into the classical and nonlocal dispersion relations, respectively. (b) Linear growth rates (left $y$ axis) and electron effective free path $\lambda_{eff}$ (right $y$ axis) for different laser intensities. The peak laser intensities are set to $1, 4, 8$ and $10\times10^{14}\ \rm{W/cm^2}$, respectively.}
\end{figure}

To understand completely the effect of the nonlocal heat flux on the perturbations with different wavelengths, we have performed simulations with wavelengths of $8, 15, 20$, and $30\ \rm{\mu m}$, respectively. All other parameters are kept the same as that in the $10\ \rm{\mu m}$ case. The results are shown in Fig. \ref{fig5}(a). The simulations of the classical SH model are in good agreement with the classical ATRI theory at all perturbation wavelengths. For the SNB model, the simulations of the long wavelengths are consistent with the classical theory. However, for the short wavelength perturbations, the classical theory, even taking the relevant hydrodynamic parameters into account, deviates from the simulations, while our nonlocal theory agrees well with the simulations. This confirms that it is necessary to consider the nonlocal heat flux appropriately in the energy equation, which can increase the cutoff wavelength of the instability. As the laser intensity increases, both the nonlocal heat flux and the electron effective free path $\lambda_{eff}$ increase, resulting in a more pronounced influence of the nonlocal effect on the growth of ARTI. We have studied the effect of the magnitude of the nonlocal heat flux on ARTI by using different laser intensities. The perturbation wavelength here is $10\ \rm{\mu m}$, and other parameters remain unchanged. As shown in Fig. \ref{fig5}(b), $\lambda_{eff}$ indeed increases as the laser intensity increases, and the classical theory gradually deviates from the simulations. This is in agreement with the experimental observations of Smalyuk et al.\cite{smalyukRayleighTaylorGrowthStabilization2008,smalyukSystematicStudyRayleigh2008}. It is seen that the theoretical model coupled with the nonlocal heat flux consists better with the simulations in all intensity ranges.

\section{Conclusions}
In summary, we have conducted a pioneering study on the impact of nonlocal electron transport on ARTI, encompassing the comprehensive processes of laser heating and transport, within a laser-irradiated target. The nonlocal effects are crucial for laser driven ICF target design, and their stabilizing effect proves advantageous in mitigating the hydrodynamic instability growth within the target. The classical theory fails to predict the ARTI growth rate in short wavelength perturbations due to ignorance of the hot electron preheating. A linear theory coupled with nonlocal heat flux is developed, and it gives consistent results to numerical simulations. The nonlocal heat flux has different degrees of suppression on different perturbation wavelengths. This suppression effect is particularly efficient for short wavelength perturbations, meaning that the nonlocal thermal conduction increases the cutoff wavelength of the perturbation.

The electron effective free path $\lambda_{eff}$ is related to the magnitude of the nonlocal heat flux and can serve as an indicator of the intensity of preheating. The results of the local linear theory applying increasing the ablation velocity and density gradient scale length gradually deviate from the simulation results when $\lambda_{eff}$ increases, while the results of the nonlocal linear theory agree with the simulation results. The suppression of perturbations by nonlocal heat flux will be enhanced when $\lambda_{eff}$ increases. The linear growth theory of ARTI and the influence of electron nonlocal transport on ARTI presented here should provide a important reference for laser and target design in laser fusion.

\section{Methods}
\subsection{Simulation Method}
The simulations in this work are carried out by using the hydrodynamics code FLASH\cite{fryxellFLASHAdaptiveMesh2000}, which includes capabilities for electron thermal conduction, laser energy deposition and multigroup radiative diffusion, and is widely used to carry out modelling of high energy density physics. We have extended the code to include electron nonlocal heat transport modelling capabilities. The classical heat conduction equation is
\begin{equation} \label{SMEQ1}
	\rho {c_v}\frac{{\partial {T}}}{{\partial t}} = \nabla  \cdot \kappa \nabla T,
\end{equation}
where $\rho$ is the mass density, $c_{v}$ is the specific heat at constant volume, $T$ is the temperature, $\kappa$ is the thermal conductivity. The SNB model\cite{schurtzNonlocalElectronConduction2000} introduces nonlocal heat flux, which modifies the heat conduction equation
\begin{equation} \label{SMEQ2}
	\rho {c_v}\frac{{\partial {T}}}{{\partial t}} = \nabla  \cdot \kappa \nabla T+\sum\limits_g {\frac{{{\lambda _g}}}{3}} \nabla {H_g},
\end{equation}
where $H_g$ is the solution of the multigroup diffusion equations,
\begin{equation}\label{SMEQ3}
	\left( {\frac{1}{{{\lambda _g}}} - \nabla \cdot \frac{{{\lambda _g}}}{3}\nabla } \right){H_g} =  \nabla  \cdot \eta _g\kappa \nabla T,
\end{equation}
${\lambda _g} = 2{\left( {{E_{g - 1/2}}/{k_B}{T}} \right)^2}{\lambda _{ei}}$ is the mean free path of electrons for different energy groups $E_g$, $E_{g-1/2}=(E_g+E_{g-1})/2$, $k_{B}$ is the Boltzmann constant, $\lambda_{ei}$ is the electron-ion mean free path, ${\eta _g} = \int_{{E_{g - 1}}/{k_B}{T}}^{{E_g}/{k_B}{T}} {{\beta ^4}{e^{ - \beta }}d\beta } /24$. The HYPRE\cite{hypre} linear algebra package is used to solve these diffusion equations.

In the simulations, we used an adaptive grid, and applied open and reflective boundary conditions in the $x$ and $y$ directions, respectively. The target uses QEOS parameters based on the Thomas-Fermi model\cite{moreNewQuotidianEquation1988}. Due to the low Z value for the CH target, we turn off the radiation unit. The initial density of the CH target is $1.1\ \rm{g/cm^3}$, and the vacuum region is filled with helium of density $10^{-6}\ \rm{g/cm^3}$. The initial temperature is uniformly set to 290 K.

\subsection{Theoretical Method}
The derivation of the ARTI linear growth rate coupling with the hot electron nonlocal transport is presented below. We consider a two-dimensional slab in the $xy$ plane, with the gravitational acceleration $g$ taken along the positive axis of $y$. According to the quasi-isobaric pressure model proposed by Kull et al.\cite{kullAblativeStabilizationIncompressible1986}, the conservation equations for mass, momentum, and energy are given by
\begin{equation} \label{Eq1}
	\frac{{\partial \rho }}{{\partial t}} + \nabla  \cdot \left( {\rho {\boldsymbol{v}}} \right) = 0,
\end{equation}
\begin{equation} \label{Eq2}
	\rho \frac{{\partial {v_x}}}{{\partial t}} + \rho {v_x}\frac{{\partial {v_x}}}{{\partial x}} + \rho {v_y}\frac{{\partial {v_x}}}{{\partial y}} =  - \frac{{\partial p}}{{\partial x}},
\end{equation}
\begin{equation} \label{Eq3}
	\rho \frac{{\partial {v_y}}}{{\partial t}} + \rho {v_x}\frac{{\partial {v_y}}}{{\partial x}} + \rho {v_y}\frac{{\partial {v_y}}}{{\partial y}} =  - \frac{{\partial p}}{{\partial y}} - \rho g,
\end{equation}
\begin{equation} \label{Eq4}
	\nabla  \cdot \left( {{\boldsymbol{v}} + \frac{{\boldsymbol{Q}}}{{{c_p}\rho T}}} \right) = 0,
\end{equation}
where $\rho$ is the mass density, ${\boldsymbol{v}}\left( {{v_x},{v_y}} \right)$ is the fluid velocity, $p$ is the pressure, $\boldsymbol{Q}$ is the heat flux, $T$ is the temperature, and $c_p$ is the specific heat capacity under constant pressure. The nonlocal multigroup diffusion equation proposed by Schurtz et al.\cite{schurtzNonlocalElectronConduction2000} is
\begin{equation} \label{Eq5}
	\sum\limits_g {\frac{{{H_g}}}{{{\lambda _g}}}}  - \nabla  \cdot \sum\limits_g {\frac{{{\lambda _g}}}{3}\nabla {H_g}}  = \nabla  \cdot \kappa \nabla T,
\end{equation}
where $\kappa$ is the thermal conductivity, $\lambda_g$ is the group mean free path, $H_g$ is the solution to the multigroup diffusion equations and the total heat flux is ${\boldsymbol{Q}} =  - \kappa \nabla T - \sum\limits_g {{\lambda _g}\nabla {H_g}} /3$. Defining ${\lambda _a} = \sum\limits_g {{H_g}} /\left( {\sum\limits_g {{H_g}/{\lambda _g}} } \right)$, ${\lambda _b} = \sum\limits_g {{\lambda _g}\nabla {H_g}} /\sum\limits_g {\nabla {H_g}} $ and ${\lambda _{eff}} = \sqrt {{\lambda _a}{\lambda _b}}$, Eq. \eqref{Eq5} can be rewritten as 
\begin{equation} \label{Eq6}
	\frac{{\sum\limits_g {{H_g}} }}{{{\lambda _a}}} - \nabla  \cdot \frac{{{\lambda _b}}}{3}\nabla \sum\limits_g {{H_g}}  = \nabla  \cdot \kappa \nabla T.
\end{equation}
All variables are assumed of the form $X\left( {x,y,t} \right) = {X_0} + \delta X\left( y \right){e^{\gamma t + ikx}}$ , where $\gamma$ is the growth rate and $k$ is the perturbation wavenumber. The linearization of Eqs. \eqref{Eq1}-\eqref{Eq4} for the perturbations $\delta X\left( y \right)$ gives
\begin{equation} \label{Eq7}
	\gamma \delta \rho  + ik{\rho _0}\delta {v_x} + ({\rho _0}\delta {v_y} + {v_0}\delta \rho )' = 0,
\end{equation}
\begin{equation} \label{Eq8}
	\gamma {\rho _0}\delta {v_x} + ik\delta p + {\rho _0}{v_0}\delta {v'_x} = 0,
\end{equation}
\begin{equation} \label{Eq9}
	\gamma ({\rho _0}\delta {v_y} + {v_0}\delta \rho ) + ik{\rho _0}{v_0}\delta {v_x} - g\delta \rho  + (2{\rho _0}{v_0}\delta {v_y} + v_0^2\delta \rho  + \delta p)' = 0,
\end{equation}
\begin{equation} \label{Eq10}
.	ik\delta {v_x} + {\left( {\delta {v_y}} \right)^\prime } - \frac{1}{{{c_p}{\rho _0}{T_0}}}\frac{{\sum\limits_g {\delta {H_g}} }}{{{\lambda _a}}} = 0,
\end{equation}
where the prime denotes differentiation with respect to $y$. The energy equation \eqref{Eq10} is obtained by the following relation
\begin{equation} \label{Eq11}
	\nabla  \cdot {\boldsymbol{Q}} =  - \frac{{\sum\limits_g {{H_g}} }}{{{\lambda _a}}}.
\end{equation}
The perturbation of the heat flux can be approximated as $\delta {\boldsymbol{Q}} =  - \kappa \nabla \left( {\delta T} \right) - {\lambda _b}\nabla \left( {\delta \sum\limits_g {{H_g}} } \right)/3$. Defining the thermal diffusivity coefficient $\chi  = \kappa /\left( {\rho {c_p}} \right)$ and the nonlocal coefficient ${\chi _n} = {\lambda _b}/\left( {3{c_p}\rho } \right)$, for the constant pressure, the following relation holds
\begin{equation} \label{Eq12}
	\frac{{\delta {\boldsymbol{Q}}}}{{{c_p}\rho T}} = \chi \nabla \frac{{\delta \rho }}{\rho } - {\chi _n}\nabla \left( {\frac{{\delta \sum\limits_g {{H_g}} }}{T}} \right).
\end{equation}
Thus Eq. \eqref{Eq10} can also be written as
\begin{equation}\label{Eq13}
	\begin{split}
	ik\delta {v_x} - {k^2}\chi \frac{{\delta \rho }}{\rho } + {k^2}{\chi _n}\left( {\frac{{\delta \sum\limits_g {{H_g}} }}{T}} \right) \\+ {\left[ {\delta {v_y} + {{\left( {\chi \frac{{\delta \rho }}{\rho }} \right)}^\prime } + {{\left( {{\chi _n}\frac{{\delta \sum\limits_g {{H_g}} }}{T}} \right)}^\prime }} \right]^\prime } = 0.
	\end{split}
\end{equation}
Taking the regions of $y < 0$ (region 1), $y = 0$, and $y > 0$ (region 2) denote heavy fluid, ablation front, and light fluid, respectively. For this sharp boundary, we integrate over the boundary to obtain the following jump conditions
\begin{equation} \label{Eq14}
	\begin{array}{l}
	\left\| {\rho \delta {v_y} + v\delta \rho } \right\| = 0,\\[2mm]
	\left\| {\delta {v_x}} \right\| = 0,\\[2mm]
	\left\| {2\rho v\delta {v_y} + {v^2}\delta \rho  + \delta p} \right\| = 0,\\[2mm]
	\left\| {\delta {v_y} + (\chi \frac{{\delta \rho }}{\rho })' - ({\chi _n}\frac{{\delta \sum\limits_g {{H_g}} }}{T})'} \right\| = 0,\\[5mm]
	\left\| {\chi \frac{{\delta \rho }}{\rho }} \right\| = 0,\\[2mm]
	\left\| {{\chi _n}\frac{{\delta \sum\limits_g {{H_g}} }}{T}} \right\| = 0,
	\end{array}
\end{equation}
$\left\| X \right\| = {X_1} - {X_2}$ denotes the jump of the variable $X$ at the interface.The perturbations on both sides of the discontinuous interface are analyzed next. Assuming that the perturbations have the form $\delta X\left( y \right) \propto {e^{ly}}$, according to Eq. \eqref{Eq6} we obtain the equation for $\delta \sum\limits_g {{H_g}}$
\begin{equation} \label{Eq15}
	\left[ {1 - \frac{{{\lambda _a}{\lambda _b}}}{3}\left( {{l^2} - {k^2}} \right)} \right]\frac{{\delta \sum\limits_g {{H_g}} }}{{{\lambda _a}}} = \kappa \left( {{l^2} - {k^2}} \right)\delta T.
\end{equation}
Assuming that the coefficient of $\delta \sum\limits_g {{H_g}}$ is not zero, the following relation holds
\begin{equation} \label{Eq16}
	\frac{1}{{{c_p}\rho T}}\frac{{\delta \sum\limits_g {{H_g}} }}{{{\lambda _a}}} = \frac{{ - \chi \left( {{l^2} - {k^2}} \right)}}{{1 - \frac{{\lambda _{eff}^2}}{3}\left( {{l^2} - {k^2}} \right)}}\frac{{\delta \rho }}{\rho }.
\end{equation}
The characteristic equation about $\delta \rho $ can be obtained from Eqs. \eqref{Eq10} and \eqref{Eq16}, while the equation about $\delta {v_y}$ can be obtained from Eqs. \eqref{Eq7} and \eqref{Eq9}.
\begin{equation} \label{Eq17}
	\left[ {\gamma  + vl - \chi \left( {{l^2} - {k^2}} \right) - \left( {\gamma  + vl} \right)\frac{{\lambda _{eff}^2}}{3}\left( {{l^2} - {k^2}} \right)} \right]\delta \rho  = 0,
\end{equation}
\begin{equation} \label{Eq18}
	\left( {{l^2} - {k^2}} \right)\left( {\gamma  + vl} \right)\rho \delta {v_y} + \left[ {l{{\left( {\gamma  + vl} \right)}^2} + g{k^2}} \right]\delta \rho  = 0.
\end{equation}
Eq. \eqref{Eq17} is a cubic equation with respect to $l$ and is no longer a simple quadratic equation (Eq. 9(b) in Ref. \onlinecite{kullAblativeStabilizationIncompressible1986}). Define $l_1$, $l_2$ and $l_3$ as solutions of Eq. \eqref{Eq17}, where $l_1 > 0 > l_2 > l_3$. For $y < 0$ , the allowed modes are $k$, $l_1$, and for $y > 0$, the allowed modes are $-k, -\gamma/v_2, l_2, l_3$, where the subscript $1$ denotes region $1$ and the subscripts $2$ and $3$ denote region $2$. For the incompressible surface modes with $l = \pm k$, it can be deduced that
\begin{equation} \label{Eq19}
	\begin{array}{l}
		\delta {v_y} = {c_{1,2}}{e^{ \pm ky}},\\
		\delta {v_x} =  \pm i\delta {v_y},\\
		\delta p =  \mp \left( {\rho /k} \right)\left( {\gamma  \pm vk} \right)\delta {v_y},\\
		\delta \rho  = 0,
	\end{array}
\end{equation}
for the convection mode with $l =  - \gamma /{v_2}$,
\begin{equation} \label{Eq20}
	\begin{array}{l}
	\delta {v_y} = {c_3}{e^{ - \gamma /{v_y}}},\\
	\delta {v_x} =  - i\gamma /\left( {kv} \right)\delta {v_y},\\
	\delta p = 0,\\
	\delta \rho  = 0,
	\end{array}
\end{equation}
and for the thermal conduction modes with $l = {l_1},{l_2},{l_3}$,
\begin{equation} \label{Eq21}
\begin{array}{l}
	\frac{{\delta \rho }}{{{\rho _0}}} = {c_{4,5,6}}{e^{{l_{1,2,3}}y}},\\
	\delta {v_y} = \frac{{{l_{1,2,3}}{{\left( {\gamma  + v{l_{1,2,3}}} \right)}^2} + g{k^2}}}{{\left( {{k^2} - l_{1,2,3}^2} \right)\left( {\gamma  + v{l_{1,2,3}}} \right)}}\frac{{\delta \rho }}{\rho }\\
	\delta {v_x} = i\frac{{k{{\left( {\gamma  + v{l_{1,2,3}}} \right)}^2} - kg{l_{1,2,3}}}}{{\left( {{k^2} - l_{1,2,3}^2} \right)\left( {\gamma  + v{l_{1,2,3}}} \right)}}\frac{{\delta \rho }}{\rho },\\
	\delta p = \left( {i/k} \right)\rho \left( {\gamma  + v{l_{1,2,3}}} \right)\delta {v_x},
\end{array}
\end{equation}
where $c_1 \sim c_6$ are arbitrary coefficients. The six jump conditions form a linear system of six unknown mode amplitudes. For $y \to {0^ - }$, the allowed modes are $k$ and $l_1$, and the perturbation is denoted as
\begin{equation} \label{Eq22}
\begin{array}{l}
	\delta {v_y} = a + {r_1}b,\\
	\delta {v_x} = ia + i{s_1}b,\\
	\delta \rho  = \frac{{{\rho _1}}}{{{v_1}}}b,\\
	\delta p =  - \frac{{{\rho _1}\gamma }}{k}\left( {1 + {u_1}} \right)a - \frac{\gamma }{k}{\rho _1}\left( {1 + {u_1}{p_1}} \right){s_1}b.
\end{array}
\end{equation}
For $y \to {0^ + }$, the allowed modes are $-k, -\gamma/v_2, l_2$ and $l_3$, and the perturbations are
\begin{equation} \label{Eq23}
\begin{array}{l}
	\delta {v_y} = c + d + {r_2}e + {r_3}f,\\
	\delta {v_x} =  - ic - id/{u_2} + i{s_2}e + i{s_3}f,\\
	\delta \rho  = \frac{{{\rho _2}}}{{{v_2}}}e + \frac{{{\rho _2}}}{{{v_2}}}f,\\
	\begin{split}
	\delta p =& \frac{{{\rho_2}\gamma }}{k}\left( {1 - {u_2}} \right)c - \frac{\gamma }{k}{\rho _2}\left( {1 + {u_2}{p_2}} \right){s_2}e \\
	&- \frac{\gamma }{k}{\rho _2}\left( {1 + {u_2}{p_3}} \right){s_3}f,
	\end{split}
\end{array}
\end{equation}
where $a\sim f$ are arbitrary constants, $u = kv/\gamma$, $p = l/k$,
\begin{equation} \label{Eq24}
	r = \frac{{l{{\left( {\gamma  + {v_0}l} \right)}^2} + g{k^2}}}{{\left( {{k^2} - {l^2}} \right)\left( {\gamma  + {v_0}l} \right)v}},
\end{equation}
\begin{equation} \label{Eq25}
	s = p\left( {1 + r} \right) + 1/u.
\end{equation}
Inserting Eqs. \eqref{Eq22} and \eqref{Eq23} into the jump conditions \eqref{Eq14} yields the following relations
\begin{equation} \label{Eq26}
	\begin{array}{l}
	R\left[ {a + \left( {1 + {r_1}} \right)b} \right] = \left[ {c + d + \left( {1 + {r_2}} \right)e + \left( {1 + {r_3}} \right)f} \right],\\
	a + {s_1}b =  - c - \frac{d}{{{u_2}}} + {s_2}e + {s_3}f,\\
	\left( {1 - \frac{1}{{{u_1}}}} \right)a + {t_1}b = \left( {1 + \frac{1}{{{u_2}}}} \right)c + 2d + {t_2}e + {t_3}f,\\
	a + {r_1}b + {p_1}{h_1}b = c + d + {r_2}e + {r_3}f + {p_2}{h_2}e + {p_3}{h_3}f,\\
	{z_1}b = {z_2}e + {z_2}f,\\
	{x_1}b = {x_2}e + {x_2}f,
	\end{array}
\end{equation}
where $R = {\rho _1}/{\rho _2}$ is the density ratio of the heavy fluid to the light fluid, $t = 1 + 2r - s\left( {p + 1/u} \right)$, $z = \chi k/v$, $x = k\gamma {\lambda ^2}\left( {1 + up} \right)/\left( {3v} \right)$, $h = x + z$. We obtain the following matrix system
\begin{equation} \label{Eq27}
	{\bf{A}} \cdot {\bf{Y}} = 0,
\end{equation}
where ${\bf{A}}$ is the coefficient matrix given by
\begin{widetext} 
	\begin{equation}\label{Eq28}
	{\bf{A}} = \left[ {\begin{array}{*{20}{c}}
			R&{R\left( {{r_1} + 1} \right)}&{ - 1}&{ - 1}&{ - \left( {{r_2} + 1} \right)}&{ - \left( {{r_3} + 1} \right)}\\
			1&{{s_1}}&1&{\frac{1}{{{u_2}}}}&{ - {s_2}}&{ - {s_3}}\\
			{1 - \frac{1}{{{u_1}}}}&{{t_1}}&{ - 1 - \frac{1}{{{u_2}}}}&{ - 2}&{ - {t_2}}&{ - {t_3}}\\
			1&{{r_1} + {p_1}{h_1}}&{ - 1}&{ - 1}&{ - \left( {{r_2} + {p_2}{h_2}} \right)}&{ - \left( {{r_3} + {p_3}{h_3}} \right)}\\
			0&{{z_1}}&0&0&{ - {z_2}}&{ - {z_2}}\\
			0&{{x_1}}&0&0&{ - {x_2}}&{ - {x_3}}
	\end{array}} \right],
	\end{equation}
\end{widetext}
and ${\bf{Y}}$ is a vector of undetermined coefficients given by
\begin{equation} \label{Eq29}
	{{\bf{Y}}^T} = \left[ {\begin{array}{*{20}{c}}
			a&b&c&d&e&f
	\end{array}} \right].
\end{equation}
The dispersion relation with respect to $\gamma$ can be derived by setting the determinant of the coefficient matrix ${\bf{A}}$ equal to zero. Consistent with Ref. \onlinecite{kullAblativeStabilizationIncompressible1986}, the dimensionless wavenumber $K = k/{k_{c1}}$, the reciprocal of Froude number $S = {k_{c1}}{\chi _1}/{v_1}$, the dimensionless growth rate ${\gamma _N} = \gamma /\left( {{k_{c1}}{v_1}} \right)$ and the dimensionless free path $\varLambda = {k_{c1}}{\lambda _1} = {k_{c1}}{\lambda _2}/{R^3}$ are introduced.



\begin{acknowledgments}\suppressfloats
This work was supported by the National Natural Science Foundation of China (Grant Nos. 12175309, 11975308, 12005297, and 12275356), the Strategic Priority Research Program of Chinese Academy of Science (Grant Nos. XDA25050200, XDA25050400, and XDA25010100), the Defense Industrial Technology Development Program (Grant. JCKYS2023212807). X.H.Y. also acknowledges the financial support from Fund for NUDT Young Innovator Awards (No. 20180104).
\end{acknowledgments}

\section*{Reference}

\end{document}